\documentclass[authoryear,final,5p,times,twocolumn]{elsarticle}

 \usepackage{graphicx}

\usepackage{amssymb}

\journal{New Astronomy}

\def\aap{A\&A}

\def\aj{AJ}
\def\apj{ApJ}

\def\apjl{ApJ}

\def\gca{Geochim. Cosmochim. Acta}
\def\mnras{MNRAS}

\def\aapr{A\&ARev}
\def\astrobj#1{#1}

\begin{document}

\begin{frontmatter}



\title{The X-ray under-luminosity of the O-type supergiants HD\,16691 and HD\,14947 revealed by XMM-Newton\tnoteref{xmm}}
\tnotetext[xmm]{Based on observations with XMM-Newton, an ESA Science Mission with instruments and contributions directly funded by ESA Member states and the USA (NASA).}

\author{M. De Becker}
\ead{debecker@astro.ulg.ac.be}



\address{Department of Astrophysics, Geophysics and Oceanography, University of Li\`ege, 17, All\'ee du 6 Ao\^ut, B5c, B-4000 Sart Tilman, Belgium}

\begin{abstract}
The members of the scarce category of Of$^+$ supergiants present properties that are intermediate between regular O-stars and Wolf-Rayet (WR) stars. Significant similarities between these transitional stars and WN-type objects are now clearly established, at least in the visible and near-infrared domains, pointing to common stellar wind properties. In this study, we report on the first dedicated X-ray observations of HD\,16691 (O4If$^+$) and HD\,14947 (O5f$^+$), revealing a soft thermal spectrum in agreement with the expected X-ray emission from a single O-type star. However, the X-ray luminosity of our targets is slightly lower than expected for single O-type stars, suggesting that the particular properties of their stellar wind has also a significant impact on the X-ray emission of these objects on the way to the WN category. We argue that the X-ray under-luminosity of HD\,16691 and HD\,14947 may be interpreted as the signature in X-rays of the intermediate stage between O and WR stars, as a consequence of enhanced wind density.
\end{abstract}


\end{frontmatter}

\section{Introduction} \label{intro}
A few O-type stars in the Galaxy are considered to be transition objects toward the WN-type, on the basis notably of similar spectral morphologies. Among these, one finds supergiants with significant emission lines in their visible and near-infrared spectrum such as OIf$^+$ stars \citep{contiliege,contitrans}. The probable presence of bubbles around some of these stars, reminiscent of the case of several Wolf-Rayet stars, has also been reported by \citet{cappaherbstmeier}. Some of these transition objects have been studied at length in the visible domain. For instance, three northern Of$^+$ supergiants, namely \astrobj{HD14947}, \astrobj{HD15570} and \astrobj{HD16691}, have been intensively monitored to investigate their line profile variability in the blue domain \citep{debeckerofplus}. The variability pattern revealed by the observations could be interpreted in the framework a large-scale co-rotating structure in the dense stellar wind of HD\,16691 (O4If$^+$), responsible for double-line shapes of the prominent emission lines, with a variability time-scale of 1.98\,d. Such a large-scale structure could be attributed to the interaction of the wind plasma with a significant magnetic field \citep{debeckerofplus}. On the other hand, recent theoretical modelling allowed to explain such line shapes on the basis of stellar rotation only \citep{hillier2012}, but it is still not clear at this stage whether stellar rotation alone could explain de variability pattern revealed by observations. Even though the variations exhibited by HD\,14947 (O5f$^+$) are less spectacular, qualitatively its behavior is very similar to that of HD\,16691. Despite their interesting properties, stars of this category lack dedicated investigations in X-rays. 

The X-ray emission from single O-type stars is mainly believed to be generated through strong shocks intrinsic to stellar winds, as a consequence of the line-driving instability \citep{FeldX}. On the other hand, the magnetic confinement of the wind leading outflows from both hemispheres to collide close to the equatorial plane may also constitute an additional source of X-rays, producing a hardening of the X-ray emission \citep{mcws1997}. In this context, only a few very early-type stars benefitted from a dedicated observation with recent X-ray observatories, such as {\it XMM-Newton} or {\it Chandra}. More specifically, peculiarities of the stellar wind of very early supergiants revealed in the optical domain suggest that the conditions prevailing in the X-ray active region of the stellar wind may also deviate from those of more 'regular' O-type stars. We therefore selected as targets, among the category of Of$^+$ supergiants that have been intensively studied in the optical, the stars presenting among the most striking features in their stellar winds, i.e. HD\,16691 and HD\,14947. 

The paper is organized as follows. The observations and data processing are decribed in Sect.\,\ref{observ}. The results of our analysis are presented and discussed in Sect.\,\ref{results}, and our concluding remarks are given in Sect.\,\ref{concl}.

\section{Observations} \label{observ}

HD\,16691 was a target of the {\it XMM-Newton} satellite \citep{xmm} during the 10th Announcement of Opportunity (AO10), with the proposal ID\,067110 (PI: M. De Becker), on 21st August, 2011, and HD\,14947 was a AO11 target of the same observatory on 20th January, 2013 (proposal ID\,069088, PI: M. De Becker). EPIC instruments were operated in Full Frame mode, and the medium filter was used to reject optical light. In both cases, about 15\,$\%$ of the exposure was affected by a rather high background level due to a soft proton flare, and that time interval was rejected for our analysis. Data were processed using the {\it XMM-Newton} Science Analysis Software (SAS) v.12.0.0 on the basis of the Observation Data Files (ODF) provided by the European Space Agency (ESA). Event lists were filtered using standard screening criteria (pattern $\leq$\,12 for MOS and pattern $\leq$\,4 for pn). We note that the aim point of the exposure was set to the position of the target in order to simultaneously obtain RGS spectra, but the low brightness of the two stars in X-rays did not allow us to perform any relevant analysis with these data. Our discussion will therefore focus on EPIC data. The observation parameters are summarized in Table\,\ref{obs}.

Spectra of HD\,16691 and HD\,14947 were extracted using a circular region centered on the target ($\alpha$:\,02$^h$\,42$^m$\,52.03$^s$; $\delta$:\,+56$^\circ$\,54'\,16.47'', and $\alpha$:\,02$^h$\,36$^m$\,46.99$^s$; $\delta$:\,+58$^\circ$\,52'\,33.12'', respectively), with a radius of 30''. The background spectrum was extracted in an annular region surrounding the source region, whose surface area was set to be equal to that of the source extraction region. We computed the response matrices using the dedicated tools available within the SAS software. We finally grouped our EPIC spectra to get at least 9 counts per energy bin.

\begin{table}
\caption{Observation parameters. The quoted exposure times (expressed in s) are the effective values, after rejection of high background level time intervals.\label{obs}}
\begin{center}
\begin{tabular}{c c c c c c}
\hline\hline
Target & JD & Rev. & \multicolumn{3}{c}{Exposure time} \\
\cline{4-6}
 & --2\,450\,000 &  & MOS1 & MOS2 & pn \\
\hline
\vspace*{-0.2cm}\\
HD\,16691 & 5794.938 & 2142 & 18640 & 19360 & 12227 \\
HD\,14947 & 6313.458 & 2402 & 18013 & 18080 & 15012 \\
\vspace*{-0.2cm}\\
\hline
\end{tabular}
\end{center}
\end{table}

\section{Results}\label{results}

Considering the results detailed by \citet{debeckerofplus} for their spectroscopic optical campaign, there is no reason a priori to consider HD\,16691 and HD\,14947 as binary stars. We note also that we did not find any hint for the presence of a companion through high angular resolution techniques in the literature. We will therefore adopt the very reasonable hypothesis that our targets are single stars throughout our analysis and our discussion of the results. As a consequence, we will not consider any additional source of X-rays originating from any putative wind-wind interaction region -- either thermal \citep{pittardparkin2010} or non-thermal \citep{debeckerreview} -- likely to contribute to the overall X-ray emission.

\begin{figure}[ht]
\begin{center}
\includegraphics[width=85mm]{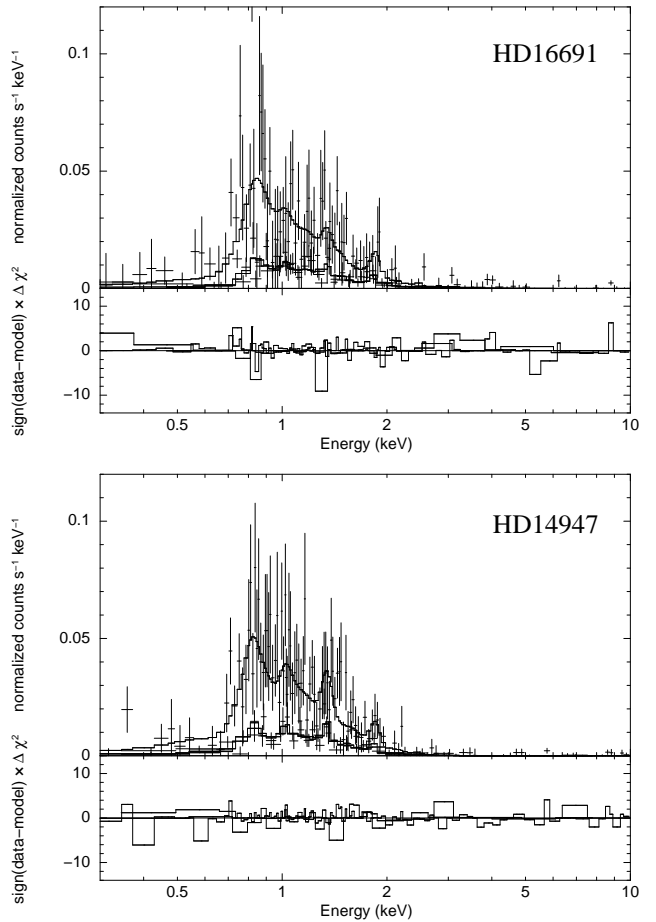}
\caption{Simultaneous fit of the three EPIC spectra with a 1-T optically thin thermal model between 0.3 and 10.0\,keV for HD\,16691 ({\it top}) and HD\,14947 ({\it bottom}). The residuals of the fit are plotted in the lower panels.\label{specfit}}
\end{center}
\end{figure}

\subsection{Spectral analysis}\label{analysis}
The X-ray spectrum of HD\,16691 is rather soft, as expected for presumably single stars. Therefore, our data do not point a priori to a hard emission component attributable to magnetic confinement emission. If such a process is at work, it should only be very weakly active. The same statement holds for the spectrum of HD\,14947. We analyzed EPIC spectra of our targets using the XSPEC software (v.12.5.1)\footnote{http://heasarc.gsfc.nasa.gov/docs/xanadu/xspec/, see also \citet{xspec1996}.}. We used composite models including absorption and emission components. For the absorption, we reproduced the photo-electric absorption by the interstellar medium (ISM) through a {\tt wabs} component, based on solar abundances from \citet{anders1982}, with the hydrogen column density (N$_\mathrm{H}$) as unique parameter. In the case of HD\,16691, we used the intrinsic (B\,--\,V)$_\circ$ (=\,--0.28) color given by \citet{martinsplez}. Taking (B\,--\,V)\,=\,0.411 as given by \citet{markova2004}, we derived E(B\,--\,V)\,=\,0.69. Inserting this value in the dust-to-gas relation given by \citet{Boh}, we derived N$_\mathrm{H}$\,=\,0.40\,$\times$\,10$^{22}$\,cm$^{-2}$. In the case of HD\,14947, the (B\,--\,V) value given by \citet{markova2004} is 0.389, and the same considerations lead to a color excess E(B\,--\,V)\,=\,0.67, converted into N$_\mathrm{H}$\,=\,0.38\,$\times$\,10$^{22}$\,cm$^{-2}$. The ISM hydrogen column density of the {\tt wabs} component was therefore fixed to those values, respectively for both stars, throughout our spectral analysis. The absorption of X-ray photons by the wind material of the target is taken into account following the same procedure as developped by \citet{nazehd108} and used in the context of other O-type star X-ray investigations \citep[e.g.][]{DeB168112}. This approach considers photo-electric absorption by an ionized material with solar abundances. In our spectral modelling procedure, that component will be referred to as {\tt wind}, with a unique column density parameters (N$_\mathrm{WIND}$) accounting for all material in the line of sight (and not only a hydrogen equivalent column as for the {\tt wabs} component used for ISM absorption).

\begin{table*}
\caption{Results of the fit of EPIC spectra between 0.3 and 10.0\,keV adopting a 1-T model. The quoted error bars are given at the 90\,$\%$ confidence level. The observed flux (f$_\mathrm{X,obs}$) and the flux corrected for the ISM absorption (f$_\mathrm{X,unabs}$) are estimated between 0.3 and 10.0\,keV. \label{bestfit}}
\begin{center}
\begin{tabular}{l c c c c}
\hline\hline
\vspace*{-0.2cm}\\
	& EPIC-MOS1 & EPIC-MOS2 & EPIC-pn & 3 EPIC instruments \\
\hline
\vspace*{-0.2cm}\\
\multicolumn{5}{l}{HD\,16691}\\
\hline
\vspace*{-0.2cm}\\
$\log$\,N$_\mathrm{WIND}$ & 21.94$_{21.76}^{22.07}$ & 21.80$_{21.51}^{22.01}$ & 21.92$_{21.74}^{22.13}$ & 21.89$_{21.80}^{21.97}$ \\
kT (keV) & 0.61$_{0.45}^{0.77}$ & 0.75$_{0.51}^{1.10}$ & 0.47$_{0.29}^{0.60}$ & 0.58$_{0.48}^{0.64}$ \\
N (cm$^{-5}$) & 3.00$_{1.61}^{8.12}$\,$\times$\,10$^{-4}$ & 1.52$_{0.00}^{3.34}$\,$\times$\,10$^{-4}$& 4.01$_{0.00}^{25.43}$\,$\times$\,10$^{-4}$ & 2.71$_{2.01}^{4.29}$\,$\times$\,10$^{-4}$ \\
$\chi^2_\nu$ (d.o.f.) & 0.844 (29) & 1.416 (26) & 0.902 (99) & 1.011 (176) \\
\vspace*{-0.2cm}\\
\hline
\vspace*{-0.2cm}\\
Count rate (cts\,s$^{-1}$) & 0.011\,$\pm$\,0.001 & 0.010\,$\pm$\,0.001 & 0.033\,$\pm$\,0.003 & -- \\
f$_\mathrm{X,obs}$ (erg\,cm$^{-2}$\,s$^{-1}$) & 6.55\,$\times$\,10$^{-14}$ & 5.70\,$\times$\,10$^{-14}$ & 5.83\,$\times$\,10$^{-14}$ & 6.04\,$\times$\,10$^{-14}$ \\
f$_\mathrm{X,unabs}$ (erg\,cm$^{-2}$\,s$^{-1}$) & 1.55\,$\times$\,10$^{-13}$ & 1.33\,$\times$\,10$^{-13}$ & 1.69\,$\times$\,10$^{-13}$ & 1.56\,$\times$\,10$^{-13}$ \\
\vspace*{-0.2cm}\\
\hline
\vspace*{-0.2cm}\\
\multicolumn{5}{l}{HD\,14947}\\
\hline
\vspace*{-0.2cm}\\
$\log$\,N$_\mathrm{WIND}$ & 21.90$_{21.70}^{22.11}$ & 22.09$_{21.88}^{22.25}$ & 21.92$_{21.80}^{22.08}$ & 21.98$_{21.90}^{22.07}$ \\
kT (keV) & 0.56$_{0.31}^{0.72}$ & 0.31$_{0.23}^{0.47}$ & 0.51$_{0.36}^{0.61}$ & 0.43$_{0.34}^{0.50}$ \\
N (cm$^{-5}$) & 2.87$_{1.44}^{16.29}$\,$\times$\,10$^{-4}$ & 16.25$_{0.00}^{76.80}$\,$\times$\,10$^{-4}$& 3.99$_{2.37}^{13.88}$\,$\times$\,10$^{-4}$ & 5.77$_{3.68}^{12.03}$\,$\times$\,10$^{-4}$ \\
$\chi^2_\nu$ (d.o.f.) & 1.137 (22) & 1.013 (26) & 0.838 (112) & 0.880 (183) \\
\vspace*{-0.2cm}\\
\hline
\vspace*{-0.2cm}\\
Count rate (cts\,s$^{-1}$) & 0.009\,$\pm$\,0.001 & 0.011\,$\pm$\,0.001 & 0.034\,$\pm$\,0.003 & -- \\
f$_\mathrm{X,obs}$ (erg\,cm$^{-2}$\,s$^{-1}$) & 5.43\,$\times$\,10$^{-14}$ & 6.10\,$\times$\,10$^{-14}$ & 6.50\,$\times$\,10$^{-14}$ & 6.15\,$\times$\,10$^{-14}$ \\
f$_\mathrm{X,unabs}$ (erg\,cm$^{-2}$\,s$^{-1}$) & 1.65\,$\times$\,10$^{-13}$ & 2.04\,$\times$\,10$^{-13}$ & 2.04\,$\times$\,10$^{-13}$ & 1.92\,$\times$\,10$^{-13}$ \\
\vspace*{-0.2cm}\\
\hline
\end{tabular}
\end{center}
\end{table*}

The X-ray emission from a single O-type star is expected to be produced by thermal processes. We therefore favored models simulating the emission from an optically thin plasma defined by a unique temperature, at ionization equilibrium ({\tt apec}). We did not allow the element abundances to deviate from the solar values given by \citet{andersgrevesse}. HD\,16691 and HD\,14947 are indeed not sufficiently evolved to present abundances altered by the central nucleosynthesis at its surface, and the nature and quality of the data do not allow to reach such levels of refinement. We fitted a 1-T model to all EPIC spectra ({\tt wabs*wind*apec}), leading to a typical plasma temperature of the order of 0.4--0.6\,keV (i.e. about 5--7\,$\times$\,10$^6$\,K). The best-fit parameters are quoted in Table\,\ref{bestfit} for all data sets, for both objects. Considering the rather low quality of the data, we will focus the discussion of the results obtained from the simultaneous fit of all three EPIC spectra. The spectra and the best-fit model are plotted in Fig.\,\ref{specfit} for both objects. We note that we also confronted our spectra to 2-T models, but it did not improve significantly the results.

\subsection{Discussion}\label{disc}

The X-ray spectrum of HD\,16691 is typical of what could be expected for a single O-type star, with a soft thermal emission spectrum characterized by a temperature of 0.4-0.6\,keV. On the basis of the unabsorbed flux (i.e corrected for the ISM absorption) quoted in Table\,\ref{bestfit}, and assusming a distance of 2.29\,kpc \citep{markova2004}, we derived an unabsorbed X-ray luminosity of 9.8\,$\times$\,10$^{31}$\,erg\,s$^{-1}$. Considering a bolometric luminosity of 8.7\,$\times$\,10$^{5}$\,L$_\odot$\,=\,3.4\,$\times$\,10$^{39}$\,erg\,s$^{-1}$ \citep{martins}, we obtained a L$_\mathrm{X,unabs}$/L$_\mathrm{bol}$ ratio of 2.9\,$\times$\,10$^{-8}$. Such a value appears quite low when compared to the expected ratio that should be close to 10$^{-7}$ \citep{sanalxlbol,naze2009} for regular single O-type stars. The X-ray luminosity is about a factor 3.5 lower than expected, which is significant regarding the rather tight dispersion of recent determinations of the L$_\mathrm{X}$--L$_\mathrm{bol}$ relation. HD\,16691 appears therefore to be significantly under-luminous in X-rays. In the case of HD\,14947, the X-ray spectrum is also quite typical of that of a single O-type star. Assuming a distance of 2.29\,kpc \citep{markova2004}, we converted the unabsorbed flux into an X-ray luminosity of 1.2\,$\times$\,10$^{32}$\,erg\,s$^{-1}$. With a bolometric luminosity of 7.4\,$\times$\,10$^{5}$\,L$_\odot$\,=\,2.9\,$\times$\,10$^{39}$\,erg\,s$^{-1}$ \citep{martins}, we determine a L$_\mathrm{X,unabs}$/L$_\mathrm{bol}$ ratio of 4.2\,$\times$\,10$^{-8}$. Here again, this value is somewhat lower ($\sim$ a factor 2.5) than expected on the basis of the L$_\mathrm{X}$--L$_\mathrm{bol}$ relation. However, one should remember that the luminosity ratios derived above are especially sensitive to the uncertainties on the distance to the stars and on the luminosity calibration proposed by \citet{martins}. To get rid of these sources of uncertainties, we also determined flux ratios, i.e. f$_\mathrm{X,unabs}$/f$_\mathrm{bol}$, for both stars. Following a standard exctinction law, we corrected visual magnitudes for the interstellar extinction on the basis of the E(B\,--\,V) values used to derive the ISM column density (Sect.\,\ref{analysis}), and we applied the bolometric correction given by \citet{martinsplez} to derive bolometric magnitudes. After converting these magnitudes into bolometric fluxes, we derived flux ratios of 8.6\,$\times$\,10$^{-8}$ and 7.9\,$\times$\,10$^{-8}$, respectively for HD\,16691 and HD\,14947. These values are slightly larger than those derived on the basis of the luminosities, but we caution that the flux-based approach is very sensitive to the effect of interstellar exctinction that is significant in the direction of our targets. In particular, the association of these stars with wind-blown bubbles \citep{cappaherbstmeier} might indeed cause some deviations with respect to the standard extinction law used in this study. In addition, the latter approach is still sensitive to uncertainties on the calibration of bolometric corrections and intrinsic colors adopted by \citet{martinsplez}, especially considering the status of these stars, with winds significantly denser than those of more classical O-type supergiants. In summary, and considering the rather tight relation derived by \cite{naze2009} (log\,(f$_\mathrm{X,unabs}$/f$_\mathrm{bol}$)\,=\,--6.45\,$\pm$\,0.51 for O-type star in the complete EPIC bandpass), we report on a slight under-luminosity of these two extreme supergiants in X-rays. Considering the rather high local column densities (N$_\mathrm{wind}$) derived from our fits, it is tempting to attribute it to enhanced absorption of X-rays by the wind material. For the sake of clarity, our luminosity and flux ratios are summarized in Table\,\ref{ratios}.

These results should be discussed in the framework of the evolution stage of our targets. As mentioned in Sect.\,\ref{intro}, HD\,16691 and HD\,14947 are considered to be in transition between the O-type and the WN-type. This statement is notably motivated by the strong similarities reported in the visible domain (strong and broad emission lines) between these objects and WN stars \citep{contiliege}. Such similarities have also been reported in the near-infrared by \citet{contitrans}. One may find it relevant to discuss the X-ray under-luminosity of HD\,16691 and HD\,14947 in the same context. Whilst the L$_\mathrm{X,unabs}$/L$_\mathrm{bol}$ is to some extent fairly well established for O-type stars, no equivalent relationship exists for Wolf-Rayet star. As already discussed for instance by \citet{stevensjenam}, the L$_\mathrm{X}$--L$_\mathrm{bol}$ relation should result from a simultaneous effect of intrinsic emission and local absorption by the stellar wind material, whose respective variations from one star to the other seem to compensate fairly well to yield a stable linear relation, at least in the case of regular O-type stars. However, the situation is quite different when WR stellar winds are concerned. It is indeed now clearly established that single Wolf-Rayet stars produce stellar winds with sufficient density and metallicity to become quite opaque to soft X-rays \citep{oskinovaWR}. A striking example is the case of the non-detection of the WN-type star \astrobj{WR40} in X-rays \citep{gossetwr40}, emphasizing the strong differences between O-type stars and their evolved WN counterparts. This critical change in the X-ray activity of early-type stars suggests that a transition should occur during the intermediate evolution stage. This is what we may be observing in the form of the X-ray under-luminosity of these two Of$^+$ supergiants. These facts are in agreement with the recent discussion by \citet{owocki2013} on the L$_\mathrm{X}$--L$_\mathrm{bol}$ relation for O-type stars. These authors indeed pointed out the expectation that moderately optically thick winds of extreme O-stars in transition to the WN type could cause some deviations with respect to the canonical L$_\mathrm{X}$--L$_\mathrm{bol}$ relation in the direction reported in the present study.

\begin{table}
\caption{Summary of luminosity and flux ratios for both stars, on the basis of X-ray fluxes corrected for ISM absorption.\label{ratios}}
\begin{center}
\begin{tabular}{c c c c c}
\hline\hline
Target &  R$_\mathrm{L}$\,=\,L$_\mathrm{X}$/L$_\mathrm{bol}$ & log\,R$_\mathrm{L}$ & R$_\mathrm{f}$\,=\,f$_\mathrm{X}$/f$_\mathrm{bol}$ & log\,R$_\mathrm{f}$ \\
\hline
\vspace*{-0.2cm}\\
HD\,16691 & 2.9\,$\times$\,10$^{-8}$ & -7.54 & 8.6\,$\times$\,10$^{-8}$ & -7.07 \\
HD\,14947 & 4.2\,$\times$\,10$^{-8}$ & -7.38 & 7.9\,$\times$\,10$^{-8}$ & -7.10 \\
\vspace*{-0.2cm}\\
\hline
\end{tabular}
\end{center}
\end{table}

As Of$^+$ stars are rather rare objects, this issue is still a matter of small number statistics. At this stage, on can also mention the case of the OIf$^+$ star VB4 in the HM1 cluster, also likely to be slightly under-luminous in X-rays with a log\,(f$_\mathrm{X}$/f$_\mathrm{bol}$) of about --7.4 (Naz\'e et al.\,2013, A\&A, submitted). The observation in X-rays of a sample of transition objects such as HD\,16691 and HD\,14947 is expected to fill a gap in the census of observational data on early-type stars. In particular, the change in the stellar wind properties occurring from the O-type to the WN-type, and already revealed in other wavebands, should be investigated in X-rays as well. The evolution of the properties of stellar winds across the evolution stages of early-type stars should alter significantly the capability of these winds to absorb X-rays after their production in deeper layers, therefore steepening significantly the L$_\mathrm{X}$--L$_\mathrm{bol}$ relation leading progressively to the complete extinction of intrinsic X-rays in many single WR stars. A large number of early-type stars, mainly in the transition zone of the evolutionary scheme, should be observed to investigate this issue.
 
\section{Conclusions}\label{concl}
We reported on the first dedicated X-ray observtion of the Of$^+$ supergiants HD\,16691 and HD\,14947, known to be in transition to the WN category. The spectrum is compatible with a thermal emission and rather soft, with a typical plasma temperature of the order of 5--7\,$\times$\,10$^{6}$\,K. However, the X-ray luminosity of HD\,16691 and HD\,14947 is quite low as compared to the emission level expected for single O-type stars. We tentatively attribute this slight -- although significant -- X-ray under-luminosity to the somewhat different properties of the stellar wind of this evolved transition object with respect to regular O-type stars. The deficit in X-rays measured in this study is indeed reminiscent of the case of single WN stars whose stellar winds are intrinsically significantly more opaque to X-ray photons than O-type winds. We argue that the X-ray under-luminosity reported here may be considered as an additional criteria supporting the idea that HD\,16691 and HD\,14947 are indeed transition objects, in addition to other criteria invoked in the optical and in the near-infrared.

\section*{Acknowledgements}
This research is supported in part through the PRODEX XMM/Integral contract. The author would like to thank Prof. G. Rauw for his comments on a preliminary version of the manuscript, and for discussions about this topic. The XMM-Newton Science Operations Center (SOC) is kindly acknowledged for the scheduling of the observations. The SIMBAD database has been consulted for the bibliography.


\end{document}